\documentclass[
reprint, amsmath,amssymb,aps,onecolumn]{revtex4-2}
\usepackage[force]{feynmp-auto}
\usepackage{dcolumn}
\usepackage{bm}
\usepackage{physics}
\usepackage[backref]{hyperref}

\begin{document}
	\preprint{APS/123-QED}
	\title{Hydrodynamics of a Multi-component Bosonic Superfluid}
	\author{Fan Zhang}
	\author{Lan Yin}
	\email{yinlan@pku.edu.cn}
	\affiliation{School of Physics, Peking University, Beijing 100871, China}
	\date{\today}
	\begin{abstract}
		In this work, we obtain the superfluid hydrodynamic equations of a multi-component Bose gas with short-ranged interactions at zero temperature under the local equilibrium assumption and show that the quantum pressure is generally present in the nonuniform case.  Our approach can be extended to systems with long-range interactions such as dipole-dipole interactions by treating the Hartree energy properly.  For a highly symmetric superfluid, we obtain the excitation spectrum and show that except for the density phonon, all other excitations are all degenerate.  The implication of our results is discussed. 
	\end{abstract}
	\maketitle

	About ninety years ago, shortly after the discovery of superfluidity \cite{KAPITZA1938Viscosity}, Tisza \cite{TISZA1938Transport} and Landau \cite{landau1941theory} proposed a two-fluid model to describe the hydrodynamics of a bosonic superfluid.  At zero temperature, the superfluid flow is described by a set of superfluid hydrodynamic equations \cite{1959Fluid}.  In this phenomenological approach based on macroscopic conservation laws \cite{1998Nonlinear,PhysRevA.72.053630,PhysRevA.95.023614,2013Lifshitz,2014Introduction} as well as the effective-theory approach \cite{svistunov}, the superfluid acceleration is determined by the local chemical potential.  For a dilute Bose gas, the superfluid dynamics are often studied by the Gross-Pitaevskii equation (GPE) \cite{gross1961structure,pitaevskii1961vortex} in which every boson is assumed is in the condensate.  The superfluid hydrodynamic equations can be derived from the GPE \cite{Stringari1996Collective,Cambridge2002Bose,Leggett,2016Bose}, but in this approach, the superfluid acceleration is not only determined by the local chemical potential, but also by the quantum pressure which depends on the density gradient.  This microscopic approach \cite{wu1996quantized,marques2000hydrodynamic,G2001Fluidity,nikuni1999two,nikuni2001landau,griffin_nikuni_zaremba_2009,A2013Collective,PhysRevA.78.043616,2010Fundamentals,1996Hydrodynamics,1997Generalized,2015Counter,PhysRevA.78.053624} is valid in the dilute region when the system is weakly interacting. The hydrodynamic equations of the bosonic superfluid also can be derived from the Schrödinger equation for short-ranged interactions by the interaction-radius expansion\cite{andreev2021hydrodynamics,andreev2021quantum}, where the quantum pressure is found in the mean-field level, but higher order terms are inconsistent with the Lee-Huang-Yang energy.  Whether or not the quantum pressure is present generally in the superfluid remains a critical issue.
	
	In this paper, we obtain the general form of the superfluid hydrodynamic equations of a multi-component bosonic system with short-ranged interactions in the path-integral formalism under the local equilibrium assumption (LEA) and shows that the quantum pressure generally exists in a nonuniform superfluid.  The quantum-pressure term in the superfluid hydrodynamic equations is necessary to recover the free-particle energy at large momentum.  Our approach can be extended to boson systems with long-ranged interactions by treating the Hartree energy as an external potential.  The superfluid hydrodynamic equations fully describe the motion of atoms in and outside the condensate in contrast to the GPE-type equations which are in terms of the condensate wavefunctions.  The Andreev-Bashkin effect of a multi-component superfluid, the drag between two components with different superfluid velocities, can be described in our approach.  We further study the case of a highly symmetric superfluid and obtain its excitation spectrum, showing that except for the density phonon, all other excitations are all degenerate.  The implication to quantum droplets is discussed.

	\textit{Superfluid Hydrodynamic Equations under LEA.}
	We first consider a multi-component bosonic system with short-ranged interactions with its action given by
	\begin{widetext}
		\begin{gather}
			S=\int dt  d \vb r \sum_{\sigma} \Psi^*_{\sigma} (\vb r,t)[i \hbar \partial_t+\frac{\hbar^2\nabla^{2}}{2m_{\sigma}}-V_\sigma(\vb r)
			] \Psi_{\sigma}(\vb r,t)+S_{int} \label{S1} , \\
			S_{int}=-\frac{1}{2}\int dt  d \vb r d \vb r' \sum_{\sigma {\sigma}'} \Psi^*_{\sigma} (\vb r,t)   \Psi^*_{\sigma'}(\vb r',t)U_{{\sigma} {\sigma}'}(\vb r-\vb r') \Psi_{\sigma'}(\vb r',t)\Psi_{\sigma}(\vb r,t)
		\end{gather}
	\end{widetext}
	where $\Psi_{\sigma}$ and $m_{\sigma}$ are the boson field and mass of the $\sigma$-component, $U_{{\sigma} {\sigma}'}(\vb r-\vb r')$ is the short-ranged interaction, and $V_\sigma(\vb r)$ is the external potential. The boson field  $\Psi_{\sigma}$ can be written in terms of the density $\rho_{\sigma}$ and phase $\Phi_{\sigma}$, $\Psi_{\sigma}=\sqrt{\rho_{\sigma}}e^{i\Phi_{\sigma}}$.
	In the superfluid phase, the density and phase fields can be separated into their mean values and fluctuations,
	\begin{eqnarray}
		\rho_{\sigma}=n_{\sigma}+\delta n_{\sigma}, \quad \Phi_{\sigma}=\phi_{\sigma}+\delta \phi_{\sigma} \label{de},
	\end{eqnarray}
	where $n_{\sigma}$ is the average density of the $\sigma$-component, $\phi_{\sigma}$ is the mean phase of the $\sigma$-component, and $\delta n_{\sigma}$ and $\delta \phi_{\sigma}$ are fluctuations in the density and phase. \cite{PopovFunctional,2016Bose,PhysRevA.66.013615,PhysRevA.95.023614}.  For a homogeneous system with translational invariance, the density $n_{\sigma}$ is the superfluid density.  When the translational invariance is broken, the superfluid density can be smaller than $n_{\sigma}$ due to Leggett's bound \cite{leggett1970can,leggett1998superfluid,chauveau2023superfluid}.  In the extreme case such as the Mott-insulator phase of a Bose gas in the optical lattice, the superfluid fraction can vanish.  In the following, we focus on the case that the external potential $V$ is not strong and slowly varies in space and time, so that the system is in the superfluid phase.  
	
	We consider the dynamical process with temporal and spatial scales of variances much larger than the intrinsic scales of the system so that the local equilibrium assumption (LEA) can be applied. In this case, the system can be divided into many small blocks in space and time with scales much smaller than those of the spatial and temporal variance and much larger than the intrinsic scales, e.g. healing lengths, interparticle distance, and the range of the interaction potential in space, local relaxation time and period of the boson de Broglie wave.  The subsystem within each block can be approximated as a macroscopic and homogeneous system in its local equilibrium, with constant densities and linear phases.  In the path integral, the contribution to the time-evolution matrix element from each block is given by
	\begin{equation}
		\mathcal{Z}_j=\int_j \mathcal{D}\Psi \mathcal{D}\Psi^* e^{\frac{i}{\hbar}S_j},
	\end{equation}
	where $j=(j_x,j_y,j_z,j_t)$ is the overall index of the 4-dimensional block, the subscript $j$ of the integral stands for the integration over the fields within the block, and $S_j$ is the action of the block.  Within the block, the average density $n_{\sigma}(\vb r, t)$ varies linearly in space and time, but this variance is very small due to the size of the block, and the average phase behaves similarly.  Thus the action of the block $S_j$ is approximately given by the action of a uniform system $S_j^U$ plus a term $S_{0j}$ containing derivatives of the average density and phase, $S_j \approx S_j^U+S_{0j}$, where
	\begin{equation}
		S_{0j} \approx - \Delta_j \sum_{\sigma}n_{j\sigma}[\hbar\partial_t\phi_{j\sigma}+\frac{\hbar^2}{2m_{\sigma}}\frac{|\nabla n_{j\sigma}|^2}{4n_{j\sigma}}] ,
	\end{equation}
	$n_{j\sigma}=n_{\sigma}(\vb r_j,t_j)$ is the mean density,  $\phi_{j\sigma}=\phi_{\sigma}(\vb r_j,t_j)$ is mean phase, $(\vb r_j,t_j)$ and $\Delta_j$ are the center and volume of the block, and $S_j^U$ is the action of the corresponding a uniform system within the same block with average density $n_{j\sigma}$ and phase $\phi_{j\sigma}$, and phase gradient $\nabla \phi_{j\sigma}$.  Thus in LEA integrating out the boson fields is equivalent to putting the time-evolution operator on the lowest-energy state of a uniform system,
	\begin{equation}
		\mathcal{Z}_j \sim e^{\frac{i}{\hbar}[S_{0j}-\Delta_j\mathcal{E}_j ]},
	\end{equation}
	where $\mathcal{E}_j$ is the energy density of the lowest-energy state of the uniform system with constant densities $n_{j\sigma}$ and phase gradient $\nabla \phi_{j\sigma}$.  After performing the integration in all the blocks, we obtain an effective action for the whole system
	\begin{equation}
		S_{eff}= \int d t \int d \vb r \big\{ -\sum_{\sigma} n_{\sigma}[\hbar \partial_t\phi_{\sigma }+V_{\sigma}
		+\frac{\hbar^2}{2m_\sigma}(\frac{|\nabla n_{\sigma}|^2}{4n_{\sigma}}+|\nabla \phi_{\sigma}|^2)]-\mathcal{E}_I\big\}, \label{S2} 
	\end{equation}
	where $\mathcal{E}_I$ is energy density due to interaction which shall be obtained from studies on uniform systems.
	
	The equation of motion can be obtained by the variational condition,
	\begin{eqnarray}
		\frac{\delta S_{eff}}{\delta \phi_{\sigma}}=0, \quad \frac{\delta S_{eff}}{\delta n_{\sigma}}=0, \label{C1}
	\end{eqnarray}
	leading to the superfluid hydrodynamic equations
	\begin{eqnarray}
		&\partial_t n_{\sigma}=-\nabla\cdot(n_{\sigma}\vb v_{\sigma}+\vb j'_{\sigma}), \label{hq1} \\ 
		&m_{\sigma}\partial_t \vb v_{\sigma}=-\nabla(\frac{1}{2}m_{\sigma} v_{\sigma}^2+\frac{\partial\mathcal{E}_I}{\partial n_{\sigma}}+V_{\sigma}-\frac{\hbar^2\nabla^2\sqrt{n_{\sigma}}}{2m_{\sigma}\sqrt{n_{\sigma}}}), \notag
	\end{eqnarray}
	where $\vb v_{\sigma}=\hbar \nabla\phi_{\sigma}/m_{\sigma}$ is the $\sigma$-component superfluid velocity, $\vb j'_{\sigma}=(j'_{\sigma x},j'_{\sigma y},j'_{\sigma z})$,  $j'_{\sigma\alpha}=\partial \mathcal{E}_I/(\hbar \partial \phi'_{\sigma\alpha}) $, and $\phi'_{\sigma\alpha}=\partial_{\alpha}  \phi_{\sigma}$, $\alpha=x,y,z$.  The anomalous current $\vb j'_\sigma$ is related to the Andreev-Bashkin effect of a multi-component superfluid, which will be discussed in the latter part of our paper.  In our approach, the superfluid hydrodynamic equations are obtained without the dilute condition and are applicable to strongly interacting bosonic systems as well as those with nonnegligible beyond-mean-field effects.  The quantum-pressure term shows up as the last term on r.h.s. of Eq.~(\ref{hq1}), consistent with the quantum potential in systems with nonuniform probability densities \cite{Bohm}.
	
	\textit{The single-component case.}
	To understand the significance of the quantum pressure, and compare the hydrodynamics equations from the GPE, we first look into the single-component case.  Due to Galilean invariance, the energy density due to interaction, $\mathcal{E}_I(n,\nabla \phi)$, is exactly the energy density of uniform ground-state $\mathcal{E}_0(n)$.  The superfluid hydrodynamic equations can be simplified to the familiar form, 
	\begin{eqnarray}
		&\partial_t n=-\nabla\cdot(n\vb v),\label{hq10}  \\ \notag
		&m\partial_t \vb v=-\nabla(\frac{1}{2}m v^2+\mu-\frac{\hbar^2\nabla^2\sqrt{n}}{2m\sqrt{n}}), 
	\end{eqnarray}
	where $\mu=V+\partial\mathcal{E}_0/\partial n$ is the local chemical potential.  If we define the order parameter $\psi=\sqrt{n}e^{i\phi}$,  it is easy to obtain its equation of motion from Eq. (\ref{hq10}), 
	\begin{eqnarray}
		i\hbar\partial_t \psi(\vb r,t)= [-\frac{\hbar^2\nabla^{2}}{2m}+\mu ]\psi(\vb r,t)\label{GP4},
	\end{eqnarray}
	which looks similar to the GPE, but it is obtained without the assumption that every atom is in the condensate.  Following a standard textbook treatment, the phonon energy of a uniform superfluid can be obtained from Eq. (\ref{hq10}), 
	\begin{equation}
		\omega_{\bf k}=\sqrt{\epsilon_{\vb k}(\epsilon_{\vb k}+2\mu)},
	\end{equation}
	where $\epsilon_{\vb k}=\hbar^2 k^2/2m $ is the kinetic energy of a free boson.  In contrast, if the quantum pressure term is missing, the phonon energy is just linearly dispersed
	\begin{equation}
		\omega'_{\bf k}=\sqrt{2\mu\epsilon_{\vb k}}.
	\end{equation}
	Thus the quantum pressure is responsible for recovering the free-particle kinetic energy at large momentum. 
	
	Although the order parameter and the condensate wavefunction have the same phase,  it relates to the total density $n$, not just the condensate density $n_0$.  In the dilute limit, to the leading order of the gas parameter, the superfluid density is equal to the condensate density, $n \approx n_{0}$, and Eq.~(\ref{GP4}) is the same as the GPE.  It is a legitimate question whether or not by modifying GPE, the higher-order effects can be captured.  To answer this question, we rewrite the continuity equation as
	\begin{equation}
		\partial_t n_{0}+\partial_t \delta n=-\nabla\cdot(\vb j_{0}+\vb \delta \vb j), \label{CEP}
	\end{equation}
	where  $\delta n= n-n_{0}$ is the quantum depletion and is a function of $n_0$ in LEA, $\vb j_{0}=n_{0} \vb v$ is the condensate current, and $\delta \vb j=\delta n\vb v$ is the current of the quantum depletion.
	To the leading order, the continuity equation holds for the condensate, $\partial_t n_{0}=-\nabla\cdot \vb j_{0}$,  which can be put into the second term on the l.h.s. of Eq.~(\ref{CEP}) for the next order, and we obtain
	\begin{equation}
		\partial_t n_{0}=-(1-\frac{d \delta n}{d n_{0}})\nabla\cdot\vb j_{0}-\nabla \frac{ \delta n}{n_{0}} \cdot \vb j_{0}. \label{CEPP}
	\end{equation}
	Eq.~(\ref{CEPP}) together with Eq.~(\ref{hq10}) cannot be rewritten in the form of a GPE-type equation,
	\begin{equation}
		i\hbar\partial_t \psi_{0}(\vb r,t)=[-\frac{\hbar^2\nabla^{2}}{2m}+\mu']\psi_{0}(\vb r,t),
	\end{equation}
	for any modified chemical potential $\mu'$.  Thus we conclude that the GPE-type equations cannot completely describe the effect of the quantum depletion.  
	
	\textit{The Andreev-Bashkin effect of a multi-component superfluid.}
	Andreev and Bashkin predicted that in a 3He-4He mixture the supercurrent of one component will drag the other component into superflow \cite{andreev1975three}.  Although it has not been observed so far, there are theoretical proposals \cite{khalatnikov1957hydrodynamics, Shevchenko2005, nespolo2017andreev} that such effect is also present in a two-component Bose superfluid.  Here we show that the Andreev-Bashkin effect of a multi-component Bose superfluid can be generally described by the superfluid hydrodynamic equations given by Eq.~(\ref{hq1}).   We consider the case with small supervelocities in the linear-response regime so that the interaction-energy density can be expanded to the quadratic order in phase gradients,
	\begin{equation}
		\mathcal{E}_I (\{n_{\sigma}\},\{\nabla\phi_{\sigma}\})\approx \mathcal{E}_0(\{n_{\sigma}\})+\frac{1}{2}\sum_{\sigma,\sigma'}\chi_{\sigma\sigma'}\nabla \phi_{\sigma}\cdot\nabla \phi_{\sigma'},
	\end{equation}
	where $\mathcal{E}_0(\{n_{\sigma}\})$ is the energy density of the uniform ground state, and  the expansion coefficient $\chi_{\sigma\sigma'}$ is related to the static current-current correlation function \cite{romito2021linear}, $\chi_{\sigma\sigma'}=\chi_{\sigma'\sigma}$.  Note that we assume the inversion symmetry which eliminates the transverse coupling between phase gradients in the quadratic order.  
	
	For the case of a uniform system with the same superfluid velocity $\vb v$ for all the components, the Galilean invariance holds
	\begin{equation}
		\mathcal{E}_I(\{n_{\sigma}\},\{\nabla\phi_{\sigma}=\frac{m_{\sigma}}{\hbar}\vb v\})= \mathcal{E}_0(\{n_{\sigma}\}),
	\end{equation}
	leading to the identity
	\begin{equation} 
		\sum_{\sigma,\sigma'}\chi_{\sigma\sigma'}\nabla \phi_{\sigma}\cdot\nabla \phi_{\sigma'} 
		=\sum_{\sigma,\sigma'} f_{\sigma\sigma'} |\vb v_{\sigma}-\vb v_{\sigma'}|^2,
	\end{equation}
	where the coefficient  $f_{\sigma\sigma'}$ depends on densities.
	In this case, the continuity equation is simply given by
	\begin{equation}
		\partial_t n_{\sigma}=-\nabla\cdot n_{\sigma}\vb v,
	\end{equation}
	which yields
	\begin{equation}
		\sum_{\sigma'} m_{\sigma'}\chi_{\sigma\sigma'}=0. \label{chi1}
	\end{equation}
	The total current is given by $\vb j=\sum_{\sigma} n_{\sigma}\vb v$, in agreement with the two-component case \cite{romito2021linear}.
	
	The superfluid hydrodynamic equations can be further written as
	\begin{subequations}
		\begin{eqnarray}
			\partial_t n_{\sigma}=-\nabla\cdot(n_{\sigma}\vb v_{\sigma}+\sum_{\sigma'} \frac{m_{\sigma'}}{\hbar^2} \chi_{\sigma \sigma'}\vb v_{\sigma'}) , \label{hl1} \\
			m_{\sigma}\partial_t \vb v_{\sigma}=-\nabla(\frac{1}{2}m_{\sigma} v_{\sigma}^2+\mu_{\sigma}-\frac{\hbar^2\nabla^2\sqrt{n_{\sigma}}}{2m_{\sigma}\sqrt{n_{\sigma}}}\label{hl2} \notag\\ 
			+\sum_{\sigma_1,\sigma_2}\frac{1}{2}(\vb v_{\sigma_1}-\vb v_{\sigma_2})^2\nabla \frac{\partial f_{\sigma_1\sigma_2}}{\partial n_{\sigma}}), 
		\end{eqnarray}\label{hl3}
	\end{subequations}
	where $\mu_{\sigma}=\partial \mathcal{E}_0/\partial n_{\sigma}+V_{\sigma}$ is the local chemical potential.  In the last term on r.h.s. of Eq.~(\ref{hl2}), the derivative $\partial f_{\sigma_1\sigma_2}/\partial n_{\sigma}$ should be significant only for $\sigma=\sigma_1$ or $\sigma=\sigma_2$, because the density change in other components barely affects the coupling between velocities of these two components.  Therefore in general the drag force is present when there is a finite velocity difference between two components.  The continuity equation (\ref{hl1}) shows that generally the current of the $\sigma$-component is different from $n_{\sigma}\vb v_{\sigma}$ due to the anomalous current, $\vb j'_\sigma=\sum_{\sigma'} m_{\sigma'} \chi_{\sigma \sigma'}\vb v_{\sigma'}/\hbar^2$.

	\textit{Boson systems with long-ranged interactions.}
	For a uniform many-body system with long-ranged interactions, the Hartree energy is always divergent in the thermodynamic limit, but this divergence is artificial for physical systems. In solids, the electrostatic energy from the ions cancels out the Hartree energy of the electrons, as demonstrated in the jellium model \cite{fetter2012quantum}.  In ultracold quantum gases, although the dipole-dipole interaction (DDI) is long-ranged, the experimental systems are always finite in size.  For these finite systems, the Hartree energy as well as the rest interaction energies can be expressed in density functionals, as in electron systems \cite{Kohn}.  In the following, by properly treating the Hartree energy, we derive the superfluid hydrodynamic equations of Bosons with long-ranged interactions within LEA.
	
	We consider a multi-component bosonic system with the action given by Eq.~(\ref{S1}), except that the interactions are now long-ranged.  We first single out the Hartree energy given by
	\begin{eqnarray}
		E_H=\frac{1}{2}\sum_{\sigma,\sigma'}\int d \vb r d \vb r'  U_{{\sigma} {\sigma}'}(\vb r-\vb r')n_{\sigma'}(\vb r') n_\sigma(\vb r)\label{EH}.
	\end{eqnarray}
	The rest steps are essentially the same as in the above section.  Under LEA, we divide the system into many small blocks.  Neighboring blocks are now coupled by the Hartree energy.  In each block, the Hartree energy is treated as a constant potential, and fluctuations can be integrated out locally.  Thus we obtain the effective action given by  
	\begin{widetext}
		\begin{eqnarray}
			S_{eff}=\int d t \int d \vb r \{ \sum_{\sigma}n_{\sigma}[ -\hbar \partial_t \phi_{\sigma}-V_\sigma -\frac{\hbar^2}{2m_{\sigma}}(\frac{|\nabla n_{\sigma}|^2}{4n_{\sigma}}+|\nabla \phi_{\sigma}|^2) +\frac{1}{2} \sum_{\sigma'}\int d \vb r' U_{{\sigma} {\sigma}'}(\vb r-\vb r')n_{\sigma'}(\vb r')]-\mathcal{E}_I\big\},\label{S2'}
		\end{eqnarray}
	\end{widetext}
	where $\mathcal{E}_I$ is now the interaction-energy density of the lowest-energy state for uniform densities and phase gradients with the Hartree energy excluded, and can be calculated in the thermodynamic limit by adding a proper background potential to cancel out the Hartree energy as in the jellium model.
	
	The superfluid hydrodynamic equations can be obtained from the equation of motion,
	\begin{eqnarray}
		&\partial_t n_{\sigma}=-\nabla\cdot(n_{\sigma}\vb v_{\sigma}+\vb j'_{\sigma}), \label{hq1'} \\ 
		&m_{\sigma}\partial_t \vb v_{\sigma}=-\nabla(\frac{1}{2}m_{\sigma} v_{\sigma}^2+\frac{\partial\mathcal{E}_I}{\partial n_{\sigma}}+V'_{\sigma}-\frac{\hbar^2\nabla^2\sqrt{n_{\sigma}}}{2m_{\sigma}\sqrt{n_{\sigma}}}),\notag 
	\end{eqnarray}
	where the modified potential is given by
	$$V'_{\sigma}(\vb r,t)=V_{\sigma}(\vb r,t)+\sum_{\sigma'}\int d \vb r'U_{{\sigma} {\sigma}'}(\vb r-\vb r')n_{\sigma'}(\vb r',t).$$
	Thus the superfluid hydrodynamic equations have the same form as those in the case with the short-ranged interactions except for the treatment of the Hartree energy.
	
	\textit{Excitations of a symmetric superfluid.}
	The dynamics of a multi-component Boson superfluid are rather complex.  In this section, we focus on a highly symmetric case where all $N$ components have the same density $n$ and mass $m$, all the intraspecies interactions are the same, and all the interspecies interactions are the same.  In the absence of superfluid currents, the excitations in a uniform system can be described by linearizing the superfluid hydrodynamic equations (\ref{hl3}),
	\begin{eqnarray}
		&\partial_t \delta n_{\sigma}=-\sum_{\sigma'}D_{\sigma\sigma'}\nabla \cdot  \delta \vb  v_{\sigma'},\label{hl4} \\ \notag
		&\partial_t \delta \vb  v_{\sigma}=-\sum_{\sigma'}[F_{\sigma\sigma'} \nabla \delta n_{\sigma'}-G_{\sigma\sigma'} \nabla^3 \delta n_{\sigma'}], 
	\end{eqnarray}
	where $\delta n_{\sigma}$ and $\delta \vb  v_{\sigma}$ are the fluctuations in superfluid density and velocity, $D_{\sigma\sigma'}=n\delta_{\sigma\sigma'}+m \chi_{\sigma\sigma'}/\hbar^2$, $F_{\sigma\sigma'}=\partial \mu_{\sigma}/m\partial n_{{\sigma}'}$, and $G_{\sigma\sigma'}=\delta_{\sigma\sigma'}\hbar^2/(4m^2n)$.  We look for plane-wave excitations with $\delta n_{\sigma}=X_{\sigma}e^{i\vb q \cdot \vb r-i\omega_q t}$ and $\delta \vb v _{\sigma}= Y_{\sigma} \vb q e^{i\vb q \cdot \vb r-i\omega_q t}$. From Eq.~(\ref{hl4}), we obtain
	\begin{eqnarray}
		&\omega_q X_{\sigma}=\sum_{\sigma'}D_{\sigma\sigma'} q^2 Y_{\sigma'}, \\ \notag
		& \omega_q Y_{\sigma}=\sum_{\sigma'}[F_{\sigma\sigma'}+G_{\sigma\sigma'} q^2] X_{\sigma'}, 
	\end{eqnarray}
	leading to the equation for $\vb X$,
	\begin{equation}
		\omega_q^2 {\vb X}=q^2 {\vb D} ({\vb F}+q^2 {\vb G}) {\vb X}.
	\end{equation}
	The excitation frequency $\omega_q$ can be solved from the Secular equation,
	\begin{equation}\label{secular}
		|q^2 {\vb D} ({\vb F}+q^2 {\vb G})- \omega_q^2 {\vb I}|=0.
	\end{equation}
	In Eq. (\ref{secular}), the matrix $\vb G$ comes from the quantum pressure is diagonal. Matrix $\vb D$, $\vb G$ and their product ${\vb D} ({\vb F}+q^2 {\vb G})$ are highly symmetric and share the same property that in each matrix all the diagonal terms are the same, and all the off-diagonal terms are the same.  From Eq. (\ref{secular}), we obtain that in the long-wavelength limit the density-phonon frequency is given by
	\begin{equation}
		\omega_{q} \approx c_p q.
	\end{equation} 
	All the other excitations are degenerate, with the excitation energy in the long-wavelength limit given by
	\begin{equation}
		\omega'_{q} \approx c_m q ,
	\end{equation}
	where $c_p^2=[(N-1)D_o+D_d][(N-1)F_o+F_d]$, $c_m^2=(D_d-D_o)(F_d-F_o)$, $D_d$ and $D_o$ are the diagonal and off-diagonal matrix-elements of $\vb D$ respectively, and $F_d$ and $F_o$ are the diagonal and off-diagonal matrix-elements of $\vb F$ respectively.  The density-phonon speed $c_p$ does not depend on the susceptibility $\chi_{\sigma \sigma'}$ based on Eq. (\ref{chi1}), consistent with the two-component case \cite{romito2021linear}.  For a highly-symmetric and weakly-interacting Bose gas, to the lowest order in the gas parameter, the susceptibility $\chi_{\sigma \sigma'}$ can be ignored, $D_o \approx 0$, $D_d \approx n$, $F_d \approx g_{1}/m$, $F_o \approx g_{2}/m$, and the phonon speeds are given by,
	\begin{eqnarray} \label{ps}
		&c_p \approx \sqrt{[g_1+(N-1)g_2]n/m},\\ \notag
		&c_m \approx \sqrt{(g_1-g_2)n/m},
	\end{eqnarray}
	where $g_1$ and $g_2$ are the intra- and inter-species coupling constants respectively.  We can infer from Eq. (\ref{ps}) that the mean-field stability condition of the superfluid state is given by $g_1>g_2$ and $g_1+(N-1)g_2>0$.  Our results can be tested if the highly symmetric Bose gas can be realized in experiments in the future.  Although for a general asymmetric Bose gas the degeneracy in the excitation spectrum is broken, its excitation spectrum can still be obtained from the superfluid hydrodynamic equations.
	
	\textit{Discussions and conclusion.}
	Quantum droplets of ultracold atoms are formed in the mean-field unstable region and stabilized by quantum fluctuations.  They are perfect examples to demonstrate the beyond-mean-field effects.  So far most experiments on quantum droplets are performed on the nonmagnetic binary boson mixture, such as the homonuclear mixture of $ ^{39}\rm{K}$\cite{Cabrera2018,Cheiney2018,semeghini2018self}, and the heteronuclear
	$ ^{39}\rm{K}$-$ ^{87}\rm{Rb}$\cite{d2019observation}, and the single-component dipolar Bose gas, such as $^{164}\rm{Dy}$\cite{Kadau2016b,Ferrier-Barbut2016,Ferrier-Barbut2016a,Schmitt2016a,Wenzel2017}, and $ ^{166}\rm{Er}$\cite{chomaz2016quantum}atoms.  In the binary boson mixture, the mean-field energy is small and attractive, and the system is stabilized by the LHY energy \cite{petrov2015}.  In the Bogoliubov theory, the LHY energy not only has a dominant real part but also a small imaginary part due to phonon instability which sparks the research on the ground state of the droplet \cite{hu2020consistent,wang2020theory,gu2020phonon}.  In a Beliav approach \cite{gu2020phonon}, it was found that higher-order fluctuations restore the phonon stability removing the imaginary part of the LHY energy, which was also confirmed in a path-integral approach \cite{2021xiong}.   For this binary droplet without phase separation \cite{he2022quantum}, the density is approximately given by the condensate, $n_\sigma \approx n_{0\sigma}$, and the interaction-energy density of the droplet is dominated by the mean-field and LHY energies \cite{petrov2015},
	\begin{equation}
		\mathcal{E}_I \approx \sum_{\sigma{\sigma}'}\frac{g_{\sigma{\sigma}'}}{2}n_{\sigma}n_{{\sigma}'}+\frac{8m^{3/2}}{15\pi^2\hbar^3}(g_{11}n_1+g_{22}n_2)^{5/2},\label{bde}
	\end{equation}
	where $m=m_1=m_2$, and $a_{\sigma\sigma'}=\frac{mg_{\sigma\sigma'}}{4\pi \hbar^2}$ is the scattering length between a $\sigma$-component atom and a $\sigma'$-component atom.  From Eq.~(\ref{GP4}), we obtain the EGPE of the two components as in Ref.~\cite{d2019observation},
	\begin{eqnarray}
		i\hbar\partial_t \psi_{1}(\vb r,t)=	[-\frac{\hbar^2\nabla^{2}}{2m}+\frac{\mathcal{\partial E}_I}{\partial n_1}]\psi_{1}(\vb r,t),\\ \notag
		i\hbar\partial_t \psi_{2}(\vb r,t)=	[-\frac{\hbar^2\nabla^{2}}{2m}+\frac{\mathcal{\partial E}_I}{\partial n_2}]\psi_{2}(\vb r,t),
	\end{eqnarray}
	where  $\psi_{\sigma}=\sqrt{n_{\sigma}}e^{i\phi_\sigma}$.     
	For the binary quantum droplet, $\delta g =g_{12}+\sqrt{g_{11}g_{22}}<0$, the densities satisfy $n_1/n_2\approx\sqrt{g_{22}/g_{11}}$ and $\frac{\mathcal{\partial E}_I}{\partial n_1}=\frac{\mathcal{\partial E}_I}{\partial n_2}=\mu$ \cite{2021xiong}, and in the solution the wavefunctions of the two components are proportional to each other.  Thus EGPE of the binary quantum droplet is reduced to a single-mode equation
	\begin{equation}
		i\hbar\partial_t \psi=[-\frac{\hbar^2\nabla^{2}}{2m}+\frac{2\sqrt{g_{11}g_{22}}\delta g|\psi|^2}{(\sqrt{g_{11}}+\sqrt{g_{22}})^2}+\frac{4m^{3/2}}{3\pi^2\hbar^3}(g_{11}g_{22})^{\frac{5}{4}}{|\psi|}^3]\psi,
	\end{equation} 
	where $\psi=\sqrt{n_{tot}}e^{i\phi}$ and $n_{tot}=n_1+n_2$.  This equation is consistent with the result in Ref.~\cite{petrov2015}.   A similar situation occurs in a single-component dipolar Bose gas, where the quantum droplet is formed in the region with the DDI strength larger than that of the repulsive $s$-wave interaction, and stabilized by the LHY energy as demonstrated from EGPE \cite{PhysRevA.93.061603,doi:10.7566/JPSJ.85.053001,PhysRevA.94.033619,baillie2018droplet}.  In the Bogoliubov theory, the LHY energy of the dipolar droplet also has an imaginary part due to phonon instability in certain propagating directions which is removed by higher-order fluctuations \cite{zhang2022phonon}.   In this case, the EGPE can be also constructed from the superfluid hydrodynamic equations.  In both cases, the EGPE is successful in the dilute region by taking into account the Lee-Huang-Yang energy which is the dominant beyond-mean-field effect.  In general, however, it is more advantageous to study the superfluid hydrodynamic equations for their broader range of validity, especially when the quantum depletion is significant.

	In conclusion, we obtain the superfluid hydrodynamic equations of a multi-component bosonic system in the path-integral formalism under the local equilibrium assumption and show that the quantum pressure term is  generally present.  Our method is valid for nonuniform and strongly interacting systems with short-ranged interactions as well as with long-ranged interactions.  The superfluid hydrodynamic equations are superior to the GPE-type equations with the condensate wavefunctions as variables, as the latter cannot fully describe the dynamics of quantum depletion.   Our approach provides a general description of the Andreev and Bashkin effect of a multi-component Bose superfluid. In the case of a symmetric superfluid, we find that except for the density phonon all other excitations are degenerate. The implications on quantum droplets are discussed.  We would like to thank Z. Q. Yu for his helpful discussions.
	
	\bibliographystyle{unsrt}
	\bibliography{hd}
\end{document}